\documentstyle[newarcrc,fleqn,epsfig]{article}

\title{Magnetic field estimation in Cyg X-3's jet}

\author{R.N. Ogley\address{Department of Physics, The Open University, Walton Hall, Milton Keynes, MK7 6AA, UK},
S.J. Bell Burnell$^{\rm a}$,
R.P. Fender\address{Astronomical Institute `Anton Pannekoek',
University of Amsterdam, 1098 SJ Amsterdam, The Netherlands},
G.G. Pooley\address{MRAO, Cavendish Laboratory, University of
Cambridge, CB3 0HE, UK},\\
E.B. Waltman\address{Remote Sensing Division, Naval Research
Laboratory, Code 7210, Washington, DC 20375-5351, USA},
M. van der Klis$^{\rm b}$}

\begin{document}
\maketitle

\begin{abstract}
Multi-wavelength photometric observations of Cygnus X-3 were carried
out at 18 cm through to 450 $\mu$m, complemented by X-ray (2-10 keV)
observations.  The system was mildly active with cm fluxes at 150 --
250 mJy.  We find the spectrum to be flat with a spectral index of
zero.  Using a modified Wolf-Rayet wind model, and assuming emission
is generated in synchrotron emitting jets from the source, we find an
upper-limit to the magnetic field of 20 G at a distance $5 \times
10^{12}$ cm is required.
\end{abstract}

\section{Introduction / History}

Cygnus X-3 is an unusual binary: a neutron star (or black hole) and
Wolf-Rayet-like secondary in a 4.8 hr period.  This system flares at
sporadic intervals, with various flux increases, but generally returns
to a quiescent level of $\sim$100 mJy.  The system has bipolar jets,
and it is generally assumed these jets are composed of
synchrotron-emitting electrons.  Because of the wind, the synchrotron
emission, which is by far the dominant emission at radio wavelengths,
becomes optically thin at increasing distance with increasing
wavelength.  The sub-mm emission is relatively unexplored; Fender {\it
et al.} [1995] first detected the emission, and found excessively high
fluxes compared with an extrapolation from the radio.  The cm
wavelength data has been analysed by Waltman {\it et al.} [1996].
Behaviour at these wavelengths is due to synchrotron emission in a
modified Wolf-Rayet wind.  We use these previous observations as a
starting point for our models.

\section{Observations}

All data were taken during 1997 September 27, MJD 50717, and
observations were arranged in order to observe the source as
close in time as possible.

Fig.~1 shows the long-term variation of Cyg X-3 from the Green Bank
Interferometer (GBI) programme, from MJD 50400 to MJD 50860 (1996
November -- 1998 February), with an insert showing the variability of
the source around the time of our SCUBA observations.  One can see
that, although a couple of major flares occurred (MJD 50485, 50610),
these have no effect on our data.  The system seems to have undergone
a minor flare event around MJD 50700, and was experiencing a state of
unrest during our observations, as it returned to quiescence.

\begin{figure}[htb]
\centering{
\epsfig{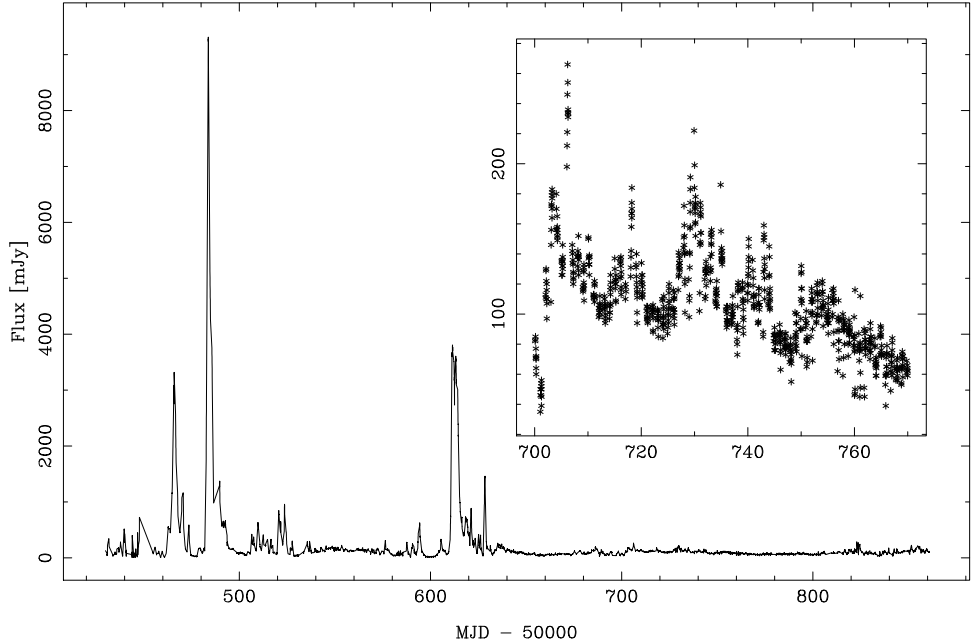}}
\caption{Flux history of Cyg X-3 from 1996 November to 1998 February
at 13 cm.  An insert from MJD 50700 to MJD 50770 is shown; our data
was taken on MJD 50717.  Although no major flares immediately precede
the observations, the system appears to be relaxing after a minor
flare -- leading to small-scale variations.}
\end{figure}

Fig.~2 shows the cm variability around the time of the SCUBA
observations.  The top panel is from the Ryle telescope at 2.0 cm and
the bottom panel is from the GBI at 3.6 cm and 13 cm.  The time of the
SCUBA observations is shown by the arrow in both plots.  Fig.~3
shows the sub-mm variability from SCUBA.  The top panel (a) is 850
$\mu$m, the bottom panel (b) is 2.0 mm.  The 450~$\mu$m datum is not
shown because poor weather prevented a reliable detection (our formal
mean for the 120 minutes of integration at this wavelength is $80 \pm
77$ mJy).

\begin{figure}[hbt]
\begin{minipage}[t]{75mm}
\epsfig{width=75mm,file=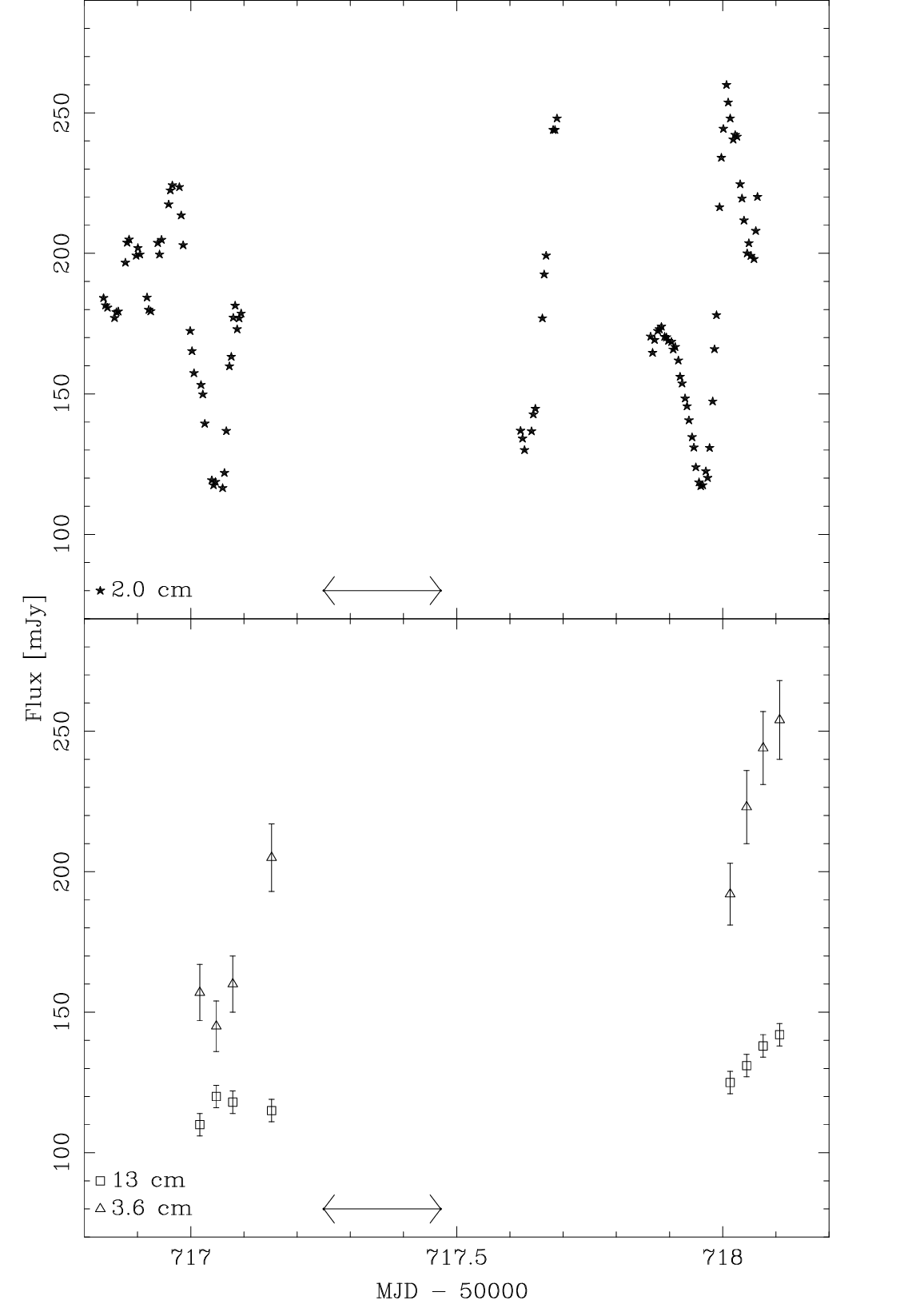}
\caption{Cyg X-3 photometry at radio wavelengths.  The top panel is
Ryle telescope data at 2.0 cm.  The bottom panel is
from the GBI, square points are at 3.6 cm and triangles are
at 13 cm.  The observations by SCUBA are shown by the
arrow.}
\end{minipage}
\hspace{\fill}
\begin{minipage}[t]{75mm}
\epsfig{width=75mm,file=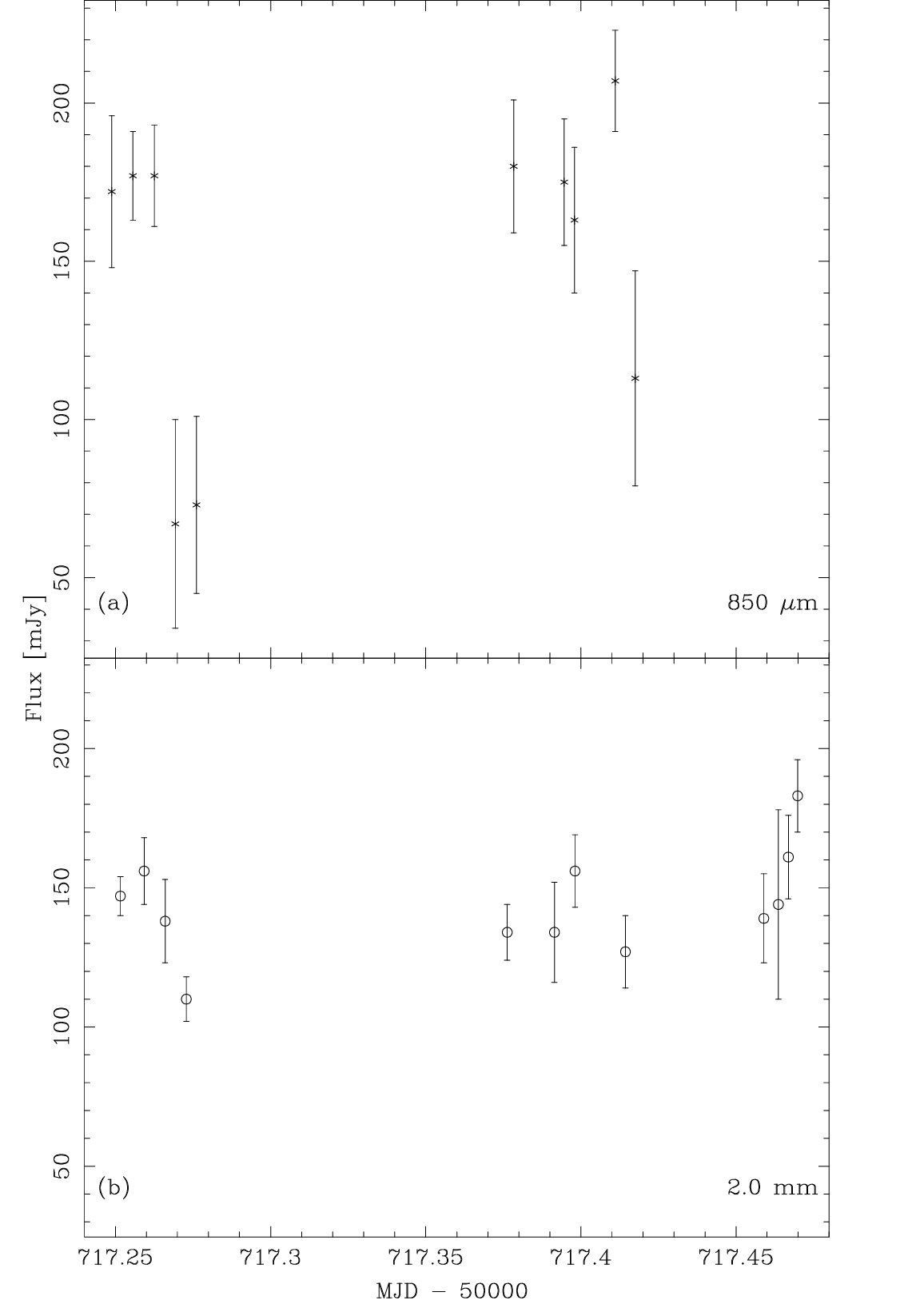}
\caption{Photometry at mm and sub-mm wavelengths.  The top panel is at
850 $\mu$m, and the bottom panel is at 2.0 mm.}
\end{minipage}
\end{figure}

Because of the variability inherent in the source, the spectrum is not
apparent.  We use the mean flux densities over the whole of the
observing times shown in figures 2 and 3 to produce an average
spectrum, detailed in table 1 and Fig.~4.  Errors are 1$\sigma$
deviations from the mean.  The datum at 450 $\mu$m is included for
reference.

\begin{table}[hbt]
\caption[]{Time-averaged flux densities}
\label{flux}
\begin{tabular*}{\hsize}{@{\extracolsep{\fill}}lllllll}
\hline
Telescope	& \multicolumn{2}{l}{Green Bank Interferometer} &
Ryle & \multicolumn{3}{l}{JCMT SCUBA}\\
Wavelength	& 13 cm & 3.6 cm & 2.0 cm & 2.0 mm & 850 $\mu$m & 450 $\mu$m \\
Flux (mJy)	& 120 $\pm$ 10 & 200 $\pm$ 40 & 186 $\pm$ 32 & 144 $\pm$ 18 & 150 $\pm$ 46 & 80 $\pm$ 77 \\
\hline
\end{tabular*}
\end{table}

\begin{figure}[htb]
\centering{\epsfig{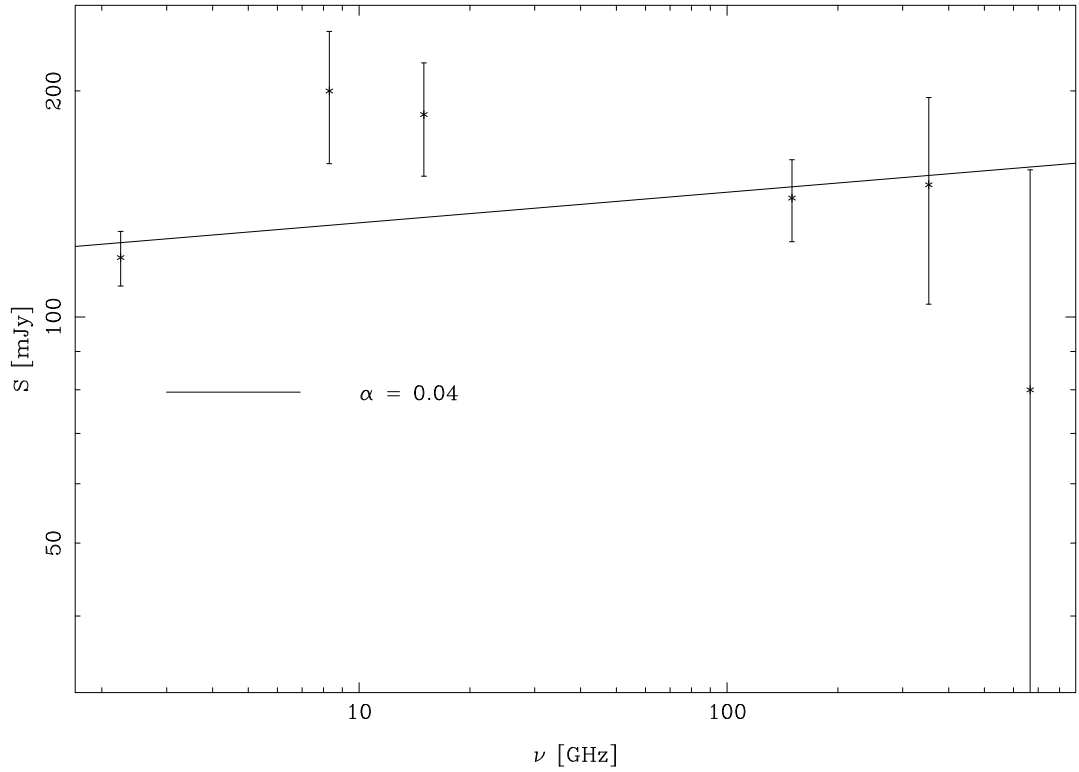}}
\caption{Spectrum from the time averaged data.  The solid line is the
best fit to all the wavelengths.}
\end{figure}

The best fit including all the data is shown by a solid line and has a
zero spectral index (within errors).  We favour a model in which the
electrons in the jet are emitting synchrotron radiation, and in which
the wind becomes optically thin to the longer wavelengths at larger
distances from the central source.  Using a stellar wind model with
the parameters given in Waltman {\it et al.} [1996], the radii at
which emission becomes optically thin are $1.5 \times 10^{14}$, $6
\times 10^{13}$, $4 \times 10^{13}$ and $5 \times 10^{12}$ cm for 13,
3.6, 2.0 cm and 850~$\mu$m respectively.  We assume material in the jet
is observed at these, and increasing, distances from the core as the
surrounding material becomes optically thin.  Electrons emitting at
$850 \mu$m would take 2000 s to travel a distance of $5 \times 10^{12}$
cm (assuming a jet velocity of 0.3 $c$), and this places a lower-limit
to their age.  Since no spectral change occurs at this wavelength, we
conclude that the high energy electrons have not aged in this time,
placing an upper-limit on the magnetic field of $\leq 20$ G at that
distance.  A higher magnetic field would create a spectral change at
the shorter wavelengths.

\end{document}